\documentstyle[prd,aps,floats,amsfonts,amssymb,amsmath]{revtex}

\begin{document}
 \renewcommand{\theequation}{\thesection.\arabic{equation}}
 \draft
  \title{Caianiello's Maximal Acceleration. Recent Developments}
\author{G. Papini$^{a,b}$\thanks{E-mail:
papini@uregina.ca}}
  \address{$^a$Department of Physics, University of Regina, Regina, Sask, S4S 0A2, Canada.}
  \address{$^b$International Institute for Advanced Scientific Studies, 89019 Vietri sul Mare (SA), Italy.}
\date{\today}
\maketitle
\begin{abstract}

A quantum mechanical upper limit on the value of particle
accelerations is consistent with the behaviour of a class of
superconductors and well known particle decay rates. It also sets
limits on the mass of the Higgs boson and affects the stability of
compact stars. In particular, type-I superconductors in static
conditions offer an example of a dynamics in which acceleration
has an upper limit.
\end{abstract}


\section{Introduction}

In 1984 Caianiello gave a direct proof that, under appropriate
conditions to be discussed below, Heisenberg uncertainty relations
place an upper limit ${\cal A}_{m}=2mc^{3}/\hbar$ on the value
that the acceleration can take along a particle worldline
\cite{cai2}. This limit, referred to as maximal acceleration (MA),
is determined by the particle's mass itself. With some
modifications \cite{wood}, Caianiello's argument is the following.

If two observables $ \hat{f}$ and $ \hat{g}$ obey the commutation relation
\begin{equation}\label{1}
\left[\hat{f},\hat{g}\right]= - i \hbar \hat{\alpha},
\end{equation}
where $ \hat{\alpha}$ is a Hermitian operator, then their uncertainties
\begin{eqnarray}\label{1bis}
 \left(\Delta f\right)^{2}&=&
<\Phi\mid\left(\hat{f}-<\hat{f}>\right)^{2}\mid\Phi>\\ \nonumber
 \left(\Delta g\right)^{2}&=&
<\Phi\mid\left(\hat{g}-<\hat{g}>\right)^{2}\mid\Phi>
\end{eqnarray}
 also satisfy the inequality
\begin{equation}\label{2}
\left(\Delta f\right)^{2}\cdot\left(\Delta g\right)^{2}\geq
\frac{\hbar^{2}}{4}<\Phi\mid\hat{\alpha}\mid\Phi>^{2},
\end{equation}
or
\begin{equation}\label{3}
\Delta f\cdot\Delta g\geq \frac{\hbar}{2} \mid<\Phi\mid\hat{\alpha}\mid\Phi>\mid .
\end{equation}
Using Dirac's analogy between the classical Poisson bracket $ \left\{f,g\right\}$ and the
quantum commutator \cite{land}
\begin{equation}\label{1ter}
\left\{f,g\right\} \rightarrow\frac{1}{i\hbar}\left[\hat{f},\hat{g}\right],
\end{equation}
one can take $ \hat{\alpha}=\left\{f,g\right\}\hat{\mathbf{1}}$.
With this substitution (\ref{1}) then yields the usual
momentum-position commutation relations. If in particular $
\hat{f}=\hat{H}$, then (\ref{1}) becomes
\begin{equation}\label{4}
\left[\hat{H},\hat{g}\right]= - i \hbar \left\{H,g\right\}\hat{\mathbf{1}},
\end{equation}
(\ref{3}) gives \cite{land}
\begin{equation}\label{5}
\Delta E\cdot \Delta g \geq \frac{\hbar}{2} \mid
\left\{H,g\right\}\mid
\end{equation}
and
\begin{equation}\label{6}
\Delta E\cdot\Delta g\geq \frac{\hbar}{2}\mid\frac{dg}{dt}\mid ,
\end{equation}
when $ \frac{\partial g}{\partial t}=0 $. Equations (\ref{5}) and
(\ref{6}) are re-statements of Ehrenfest theorem. Criteria for its
validity are discussed at length in the literature
\cite{land,mess,bal}. If $g \equiv v(t)$ is the velocity
expectation value of a particle whose energy is $ E=m c^{2}\gamma$
and it is assumed with Caianiello \cite{cai2} that $ \Delta E \leq
E$, then (\ref{6}) gives
\begin{equation}\label{7}
\mid \frac{dv}{dt}\mid \leq \frac{2}{\hbar} mc^{2}\gamma \Delta v(t) ,
\end{equation}
 where $ \gamma =\left(1 -
\frac{v^{2}}{c^{2}}\right)^{-\frac{1}{2}}$. In general and with
rigour
\begin{equation}\label{*}
 \Delta v =
\left(<v^{2}>-<v>^{2}\right)^{\frac{1}{2}}\leq v_{max}\leq c .
\end{equation}
 An essential point of Caianiello's argument is that the acceleration is largest in the rest frame
 of the particle. This follows from the transformations that
link the three-acceleration $ \vec{a}_{p}$ in the instantaneous
rest frame of the particle to the particle's acceleration $
\vec{a}' $ in another frame with instantaneous velocity $ \vec{v}$
\cite{steph}
\begin{equation}\label{1quater}
\vec{a}'=
\frac{1}{\gamma^{2}}\left[\vec{a}_{p}-\frac{\left(1-\gamma\right)\left(\vec{v}\cdot
\vec{a}_{p}\right)\vec{v}}{v^{2}}-\frac{\gamma \left(\vec{v}\cdot
\vec{a}_{p}\right)\vec{v}}{c^{2}}\right] .
\end{equation}
 The equation
\begin{equation}\label{1quinquies}
a'^{2}= \frac{1}{\gamma^{4}}\left(a_{p}^{2}-\frac{\left(\vec{a}_{p}\cdot
\vec{v}\right)^{2}}{c^{2}}\right),
\end{equation}
where $ a^{2} \equiv \vec{a}\cdot \vec{a}$, follows from
(\ref{1quater}) and shows that $ a' \leq a_{p}$ for all $
\vec{v}\neq 0$ and that $ a' \rightarrow 0 $ as $ \mid
\vec{a}_{p}\cdot \vec{v}\mid \rightarrow a_{p}c $. In addition, in
the instantaneous rest frame of the particle, $  E \leq mc^{2}$
and $ \Delta E \leq m c^{2}$ if negative rest energies must be
avoided as nonphysical. Then (\ref{7}) gives
\begin{equation}\label{8}
\mid \frac{dv}{dt}\mid \leq 2\frac{mc^{3}}{\hbar}\equiv {\cal
A}_{m} .
\end{equation}

 It is at times argued that the uncertainty relation
 \begin{equation} \label{yy}
 \Delta E \cdot
 \Delta t\geq \hbar/2
 \end{equation}
  implies that, given a fixed average energy E, a state can be
 constructed with arbitrarily large $\Delta E$, contrary to Caianiello's assumption that $\Delta E \leq E$. This
 conclusion is erroneous. The correct interpretation of (\ref{yy}) is that a quantum state
 with spread in energy $ \Delta E$ takes a time $ \Delta t \geq  \frac{\hbar}{2 \Delta E}$
 to evolve to a distinguishable (orthogonal) state. This evolution time has a lower bound.
 Margolus and Levitin have in fact shown \cite{marg} that the evolution time of a quantum
 system with fixed average energy $E$ must satisfy the more stringent limit
 \begin{equation} \label{&&&}
 \Delta t\geq \frac{\hbar}{2E} ,
\end{equation}
which determines a maximum speed of orthogonality evolution
\cite{beck,an}.
 Obviously, both
limits (\ref{yy}) and (\ref{&&&}) can be achieved only for $
\Delta E = E$, while spreads $ \Delta E > E $, that would make $
\Delta t $ smaller, are precluded by (\ref{&&&}). This effectively
 restricts $\Delta E$ to values
 $\Delta E \leq E$, as conjectured by Caianiello \cite{pap02}.

 Known
transformations now ensure that the limit (\ref{8}) remains
unchanged. It follows, in fact, that in the rest frame of the
particle the absolute value of the proper acceleration is
\begin{equation}\label{9}
\left(\mid
\frac{d^{2}x^{\mu}}{ds^{2}}\frac{d^{2}x_{\mu}}{ds^{2}}\mid\right)^{\frac{1}{2}}=
\left(\mid\frac{1}{c^{4}}\frac{d^{2}x^{i}}{dt^{2}}\mid
\right)^{\frac{1}{2}}\leq \frac{{\cal A}_{m}}{c^{2}} .
\end{equation}
Equation (\ref{9}) is a Lorentz invariant. The validity of
(\ref{9}) under Lorentz transformations is therefore assured.

The uncertainty relation (\ref{yy}) can then be used to extend
(\ref{8}) to include the average length of the acceleration $
<a>$. If, in fact, $ v(t)$ is differentiable, then fluctuations
about its mean are given by
\begin{equation}\label{del}
\Delta v \equiv v-<v>\simeq \left(\frac{dv}{dt}\right)_{0}\Delta
t+\left(\frac{d^{2}v}{dt^{2}}\right)_{0}\left(\Delta
t\right)^{2}+...  .
\end{equation}
Equation (\ref{del}) reduces to $ \Delta v\simeq
\mid\frac{dv}{dt}\mid \Delta t =<a> \Delta t$ for sufficiently
small values of $\Delta t$, or when $\mid\frac{dv}{dt}\mid$
remains constant over $\Delta t$. The inequalities (\ref{yy}) and
(\ref{*}) then yield
\begin{equation}\label{ga}
<a> \leq \frac{2c \Delta E}{\hbar}
\end{equation}
and again (\ref{8}) follows.

The notion of MA delves into a number of issues and is connected to the extended nature
of particles. In fact, the inconsistency of the point particle concept for a relativistic
quantum particle is discussed by Hegerfeldt \cite{heg} who shows that the localization of
the particle at a given point at a given time conflicts with causality.

Classical and quantum arguments supporting the existence of MA have been frequently
discussed in the literature
\cite{mthw,das,gasp1,toller,paren,vora,mash,venzo,falla,pati}. MA would eliminate
divergence difficulties affecting the mathematical foundations of quantum field theory
\cite{nester}. It would also free black hole entropy of ultraviolet divergences
\cite{hoof,suss,mcgui}. MA plays a fundamental role in Caianiello's geometrical
formulations of quantum mechanics \cite{cai4} and in the context of Weyl space
\cite{pap}. A limit on the acceleration also occurs in string theory. Here the upper
limit appears in the guise of Jeans-like instabilities that develop when the acceleration
induced by the background gravitational field is larger than a critical value for which
the string extremities become casually disconnected \cite{sanchez,gasp2,gasp3}. Frolov
and Sanchez \cite{fro} have also found that a universal critical acceleration must be a
general property of strings.

Incorporating MA in a theory that takes into account the limits
(\ref{8}) or (\ref{9}) in a meaningful way from inception is a
very important question. In Caianiello's reasoning
     the usual Minkowski line element
\begin{equation}\label{mink}
   ds^2=\eta_{\mu\nu}dx^{\mu}dx^\nu,
   \end{equation}
   must be replaced with the
   infinitesimal element of distance in the eight-dimensional
   space-time tangent bundle $ TM $
   \begin{equation}\label{3.1}
   d\tau^2=\eta_{AB}dX^AdX^B ,
   \end{equation}
   where $  A, B = 0, \ldots, 7,
    \eta_{AB}=\eta_{\mu\nu}\otimes \eta_{\mu\nu},
      X^{A}=\left(x^{\mu}\,\frac{c^2}{{\cal A}_m}
   \frac{dx^{\mu}}{ds}\right),
   x^{\mu}=(ct,\vec{x})$ and $ dx^{\mu}/
   ds\equiv\dot x^{\mu}$ is the four-velocity.
   The invariant line element (\ref{3.1}) can therefore be written
   in the form
   \begin{equation}\label{3.2}
   d\tau^2=\eta_{\mu\nu}dx^{\mu}dx^\nu +{1\over {\cal A}_m^2}
   \eta_{\mu\nu}d\dot x^{\mu}d\dot x^\nu=
    \left[1+\frac{\ddot x_\mu\ddot x^\mu}{{\cal A}_m^2}\right]
   \eta_{\mu\nu} dx^\mu dx^\nu ,
   \end{equation}
   where all proper accelerations are normalized by ${\cal A}_m $.
The effective space-time geometry experienced by accelerated test
particles contains therefore mass-dependent corrections which in
general induce curvature and give rise to a mass dependent
violation of the equivalence principle. In the presence of
gravity, we replace $\eta_{\mu\nu}$ with the corresponding metric
tensor $g_{\mu\nu}$, a  natural choice which preserves the full
structure introduced in the case of flat space. In the classical
limit $\left ({\cal A}_m\right)^{-1} = {\hbar\over km
c^3}\rightarrow 0$ the terms contributing to the modification of
the geometry vanish and one returns to the ordinary space-time
geometry.

The model of Ref.\cite{cai4} has led to interesting results that
range from particle physics to astrophysics and cosmology
\cite{pap1,cai5}.

A second, equally fundamental problem stems from Caianiello's
original paper "Is there a maximal acceleration?"\cite{cai1}. In
particular, is it possible to find physical conditions likely to
lead to a MA? An approach to this problem is illustrated below by
means of three examples that are offered here as a tribute to the
memory of the great master and to the ever challenging vitality of
his thoughts.

The limits (\ref{8}) and (\ref{9}) are very high for most
particles (for an electron $ {\cal A}_{m}\sim 4.7\times 10^{31}
cm\, s^{-2}$) and likely to occur only in exceptional physical
circumstances. The examples considered in this work involve
superfluids, type-I superconductors in particular, that are
intrinsically non-relativistic quantum systems in which $ \Delta
E$ and $\Delta v$ can be lower than the uncertainties leading to
the limit (\ref{8}). Use of (\ref{6}) is here warranted because
the de Broglie wavelengths of the superfluid particles vary little
over distances of the same order of magnitude \cite{land,bal}.
Superfluids are particular types of quantum systems whose
existence is predicated upon the formation of fermion pairs. They
behave, in a sense, as universes in themselves until the
conditions for pairing are satisfied. The dynamics of the
resulting bosons differs in essential ways from that of "normal"
particles. The upper limits of the corresponding dynamical
variables like velocity and acceleration must first of all be
compatible with pairing and and may in principle differ from $ c$
and $ {\cal A}_{m}$. If the limit on $ v$ is lower, then according
to (\ref{6})the possibility of observing the effects od MA is
greater. It is shown below that indeed type-I superconductors in
static conditions offer an example of a dynamics with a MA.

Lepton-lepton interactions with the final production of gauge
bosons also seem appropriate choices because of the high energies,
and presumably high accelerations, that are normally reached in
the laboratory. The application of (\ref{6}) is justified by the
high momenta of the leptons that satisfy the inequality $\Delta
x\Delta p\gg \hbar$ \cite{land,bal}.

A last example regards the highly unusual conditions of matter in the interior of white
dwarfs and neutron stars. In this case the legitimacy of (\ref{6}) is assured by the
inequality $ \frac{N}{V} \lambda_{T}^{3}\ll 1$, where $ \frac{N}{V}$ is the particle
density in the star and $ \lambda_{T}= \frac{2\pi\hbar^{2}}{mkT}$ is of the same order of
magnitude of the de Broglie wavelength of a particle with a kinetic energy of the order
of $ kT$ \cite{land,bal}.

\section{Type-I Superconductors}

The static behavior of superconductors of the first kind is adequately described by
London's theory \cite{degennes}. The fields and currents involved are weak and vary
slowly in space. The equations of motion of the superelectrons are in this case
\cite{til}
\begin{eqnarray}\label{10}
\frac{D\vec{v}}{Dt}&=&\frac{e}{m}\left[\vec{E}+ \left(\frac{\vec{v}}{c}\times
\vec{B}\right)\right]\nonumber\\ &=& \frac{\partial \vec{v}}{\partial
t}+\left(\vec{v}\cdot \vec{\nabla}\vec{v}\right) .
\end{eqnarray}
On applying (\ref{6}) to (\ref{10}), one finds
\begin{equation}\label{11}
\sqrt{\left(\frac{1}{2}\vec{\nabla}v^{2}-\vec{v}\times \left(\vec{\nabla}\times
\vec{v}\right)\right)^{2}}\leq \frac{2}{\hbar}\bigtriangleup E\cdot \bigtriangleup v  ,
\end{equation}
and again
\begin{widetext}
\begin{equation}\label{12}
\sqrt{\frac{1}{4}\left(\nabla_{i}v^{2}\right)^{2}+\frac{e}{m
c}\epsilon_{ijk}\left(\nabla^{i}v^{2}\right)v^{j}B^{k}+ \left(\frac{e}{m
c}\right)^{2}\left[v^{2}B^{2}-\left(v_{i}B^{i}\right)^{2}\right]} \leq
\frac{2}{\hbar}\triangle E \bigtriangleup v  ,
\end{equation}
\end{widetext}
where use has been made of London's equation
\begin{equation}\label{13}
\vec{\nabla}\times \vec{v}=-\frac{e}{m c}\vec{B} ,
\end{equation}
and $ \epsilon_{ijk}$ is the Levi-Civita tensor. Static conditions, $ \frac{\partial
v}{\partial t}=0 $, make (\ref{11}) and (\ref{12}) simpler. It is also useful, for the
sake of numerical comparisons, to apply (\ref{12}) to the case of a sphere of radius $ R$
in an external magnetic field of magnitude $ B_{0}$ parallel to the polar axis. This
problem has an obvious symmetry and can be solved exactly. The exact solutions of
London's equations for $ r\leq R $ are well-known \cite{london} and are reported here for
completeness. They are
\begin{equation}\label{14}
B_{r}=\frac{4 \pi}{\beta^{2}c}\frac{1}{r}\frac{1}{\sin
\theta}\frac{\partial}{\partial\theta}\left(\sin\theta j_{\varphi}\right)
\end{equation}
\begin{equation}\label{15}
B_{\theta}=-\frac{4 \pi}{\beta^{2}c}\frac{1}{r}\frac{\partial}{\partial r}\left(r
j_{\varphi}\right) ,
\end{equation}
\begin{equation}\label{16}
j_{\varphi}=nev_{\varphi}=\frac{A}{r^{2}}\left(\sinh \beta r - \beta r \cosh \beta
r\right)\sin\theta ,
\end{equation}
where $A=-\frac{c}{4\pi}\frac{3B_{0}}{2}\frac{R}{\sinh\beta r}$, $
n$ is the density of superelectrons and $\beta=\left(\frac{4\pi n
e^{2}}{m c^{2}}\right)^{\frac{1}{2}}$ represents the reciprocal of
the penetration length. On using (\ref{10}), the inequality
(\ref{11}) can be transformed into
\begin{equation}\label{17}
\mid E_{r}\mid \leq \frac{\mid
v_{\varphi}B_{\theta}\mid}{c}+\sqrt{\left(\frac{2mc}{e\hbar}\right)^{2}\left(\Delta
E\right)^{2}\left(\Delta v\right)^{2}-\left(\frac{v_{\varphi}}{c}B_{r}\right)^{2}} .
\end{equation}
For a gas of fermions in thermal equilibrium $ \Delta E\sim \frac{3}{5}\mu$, $ \Delta
v\sim \frac{3}{2}\sqrt{\frac{\mu}{2m}}$ and the chemical potential behaves as $ \mu
\approx \epsilon_{F}-\frac{\left(\pi kT\right)^{2}}{12\epsilon_{F}}\approx
\epsilon_{F}\sim 4.5\times 10^{-12}erg $ for $ T $ close to the transition temperatures
of type-I superconductors. The reality of (\ref{17}) requires that $ \Delta E\geq
\mu_{B}B_{r}$, where $ \mu_{B}= \frac{e\hbar}{2mc}$ is the Bohr magneton, or that $
\frac{3}{5}\epsilon_{F}\geq \mu_{B}B_{r}$. This condition is certainly satisfied for
values of $ B_{r}\leq B_{c}$, where $ B_{c} $ is the critical value of the magnetic field
applied to the superconductor. From (\ref{17}) one also obtains
\begin{equation}\label{18}
\mid E_{r}\mid\leq
\frac{3}{2m}\left(\frac{\epsilon_{F}}{2}\right)^{\frac{1}{2}}\left[\frac{\mid
B_{\theta}\mid}{c}+\sqrt{\left(\frac{3\epsilon_{F}}{5\mu_{B}}\right)^{2}-\left(\frac{B_{r}}{c}
\right)^{2}}\right],
\end{equation}
which is verified by the experimental work of Bok and Klein
\cite{bok}. More restrictive values for $ \Delta E$ and $ \Delta
v$ can be obtained from $ B_{c} $ . The highest value of the
velocity of the superelectrons must, in fact, be compatible with $
B_{c} $ itself, lest the superconductor revert to the normal
state. This value is approximately a factor $ 10^{3}$ smaller than
that obtained by statistical analysis. The upper value $ v_{0}$ of
$ v_{\varphi}$ is at the surface. From $ \Delta E \leq
\frac{1}{2}m v_{0}^{2}, \Delta v \leq v_{0}$ and (\ref{17}) one
finds that at the equator, where $ B_{r}=0$, $ E_{r} $ satisfies
the inequality
\begin{equation}\label{19}
\mid E_{r}\mid \leq \frac{v_{0}}{c}\left(\mid B_{\theta}\mid
+\frac{v_{0}^{2}}{2\mu_{B}}\right).
\end{equation}
For a sphere of radius $R=1 cm$ one finds $v_{0}\simeq 4.4\times
10^{4} cm\,s^{-1}$  and $ E_{r} \leq 69\, N/C$. If no magnetic
field is present, then (\ref{19}) gives $ E_{r}\leq 4.2\, N/C $.
On the other hand London's equation gives
\begin{equation}\label{**}
E_{r}=\frac{m}{2e}\frac{\partial v_{\varphi}^{2}}{\partial
r}\simeq 0.32\, N/C.
\end{equation}
The inequality (\ref{**}) agrees with the experimental data
\cite{bok}. The MA limits (\ref{18}) and (\ref{19}) are therefore
consistent with (\ref{**}) and its experimental verification.

\section{High energy lepton-lepton interactions}

Consider the process $ A +B \rightarrow D $ by which two particles
$ A$ and $ B$ of identical mass $ m$ give rise to particle $ D$. $
A$, or $ B$, or both particles may also be charged. Assume
moreover that $ D $ is produced at rest in its proper frame. The
width of $ D$ is $ \Gamma (AB)$ and the time during which the
process takes place in the center of mass frame of $ A $ and $ B$
is $ \Delta t \simeq \frac{\hbar}{\Gamma(AB)}$. The acceleration
of $ A$ and $ B$ is thought to remain close to zero until $ D$ is
produced. If this were not the case, radiation would be produced
and the process would acquire different characteristics. The root
mean square acceleration of the reduced mass $ m_{r}$ over $
\Delta t $ will in general be $ a_{r}\simeq \frac{c}{\gamma \Delta
t}$, where $ \gamma $ refers to
 the velocity $ v $ of $ m_{r}$ \cite{weber}.
The MA limit (\ref{9}), applied to $ a_{r}$, gives $ \frac{c}{\gamma\Delta t}\leq
\frac{m_{D}c^{3}}{\hbar}$ , or
\begin{equation} \label{20}
\Gamma(AB) \leq m c^{2}\gamma ,
\end{equation}
and $ v$ obeys the condition $ 2 mc^{2}\gamma =m_{D} c^{2}$, or $
\gamma = \frac{m_{D}}{2m}$ .

The limit (\ref{20}) can also be written in the form
\begin{equation} \label{21}
\Gamma(AB)=\frac{p_{f}}{32 \pi^{2}m_{D}}\int |M(AB)|^{2}\leq
\frac{m_{D}c^{2}}{2},
\end{equation}
where $ M(AB)$ is the invariant matrix element for the process.
Several processes can now be considered. For the process $
e^{+}+e^{-}\rightarrow Z^{0}$ one has (in units $ \hbar = c = 1 $)
$ \Gamma (e^{+}e^{-}\rightarrow Z^{0})\simeq 0.08391\, GeV $, $
m_{Z_{0}}=91.188\, GeV $ and the inequality (\ref{21}) is
certainly satisfied.

Similarly, $ \Gamma( W\rightarrow e\nu_{e})\simeq 0.22599\, GeV
<\frac{80.419}{2}GeV$.

One expects the limit (\ref{21}) to be less restrictive at lower
energies. In the case of $ e^{+}+e^{-}\rightarrow J/\psi $ one in
fact finds $ \Gamma_{ee \rightarrow J/\psi}\simeq 5.2\, KeV $,
while the r.h.s. of (\ref{21}) gives $ \frac{3096.87}{2} MeV $.

Alternatively, (\ref{21}) can be used to find an upper limit to the value of $ m_{D}$.
These values depend, of course, on theoretical estimates of the corresponding decay
widths \cite{renton}. From
\begin{equation}\label{22}
\Gamma(Z^{0}\rightarrow ee)\simeq \frac{G_{F}m_{Z{0}}^{3}}{12 \pi \sqrt{2}}\leq
\frac{m_{Z^{0}}}{2},
\end{equation}
one obtains $ m_{Z^{0}}\leq(\frac{6\pi\sqrt{2}}{G_{F}}
)^{\frac{1}{2}}\simeq 1512\,GeV$ which is a factor $ 16.6$ larger
than the experimental value of the mass of $Z^{0}$. Analogously
\begin{equation}\label{23}
\Gamma(W\rightarrow d\bar{u})\simeq
\frac{G_{F}4.15m_{W}^{3}}{6\sqrt{2}\pi}\leq\frac{m_{W}}{2}
\end{equation}
yields $m_{W}\leq 525\, GeV$ which is $ \sim 6.53$ times larger
than the experimental value of the boson mass. Finally, from
\begin{equation}\label{24}
\Gamma(J/\psi\rightarrow ee)\simeq \frac{16\pi
\alpha^{2}0.018}{m_{J/\psi}}\leq\frac{m_{J/\psi}}{2} ,
\end{equation}
one finds $m_{J/\pi}\geq4.6\times10^{-2}\,GeV$, a lower limit
which is $1.5\times10^{-2}$ times smaller than the known value of
$m_{J/\psi}$.

For the Higgs boson the inequality
\begin{equation}\label{25}
\Gamma(H^{0}\rightarrow ee)\simeq
\frac{G_{F}m_{e}^{2}m_{H}}{4\sqrt{2}\pi}\left(1-\frac{4m_{e}^{2}}{m_{H}^{2}}\right)^{\frac{3}{2}}
\leq\frac{m_{H}}{2}
\end{equation}
is always satisfied for $m_{H}\geq 2m_{e}$. Finally, from \cite{renton}
\begin{widetext}
\begin{equation}\label{26}
\Gamma(H^{0}\rightarrow ZZ)\simeq
\frac{G_{F}m_{Z}^{2}m_{H}}{16\pi\sqrt{2}x_{Z}}\left(1-x_{Z}\right)^{\frac{1}{2}}
\left(3x_{Z}^{2}-4x_{Z}+4\right)\leq \frac{m_{H}}{2},
\end{equation}
\end{widetext}
where $x_{Z}\equiv\frac{4m_{Z}^{2}}{m_{H}^{2}}$, one finds
\begin{equation}\label{00}
2m_{Z}\leq m_{H}\leq 1760\,GeV .
\end{equation}
Both lower and upper limits are compatible with the results of
experimental searches. The upper limit also agrees with Kuwata's
analogous attempt \cite{kuwata} to derive a bound on $ m_{H}$ from
the MA constraint, when account is taken of the fact that in
Ref.\cite{kuwata} MA is $ {\cal A}_{m}/2 $ and therefore $
m_{H}\leq 500\, GeV $.

\section{White dwarfs and neutron stars}

 The standard expression for the ground state energy of an ideal fermion gas
inside a star of radius $R$ and volume $V$ is
\begin{equation}\label{27}
E_{0}(r)= \frac{m^{4}c^{5}}{\pi^{2}\hbar^{3}}V f(x_{F}),
\end{equation}
where $ x_{F}\equiv \frac{p_{F}}{mc}$, $ p_{F}\equiv
\left(3\pi^{2}\frac{N}{V}\right)^{\frac{1}{3}}\hbar$ is the Fermi momentum, $ 0\leq r\leq
R$, $N$ is the total number of fermions and
\begin{equation}\label{28}
f( x_{F}) = \int_{0}^{x_{F}}dx x^{2}\sqrt{1 + x^{2}}.
\end{equation}
The integral in (\ref{28}) is approximated by
\begin{equation}\label{29}
f(x_{F})\simeq \frac{1}{3}x_{F}^{3}\left(1 + \frac{3}{10}x_{F}^{2}+...\right),  x_{F}\ll
1
\end{equation}
in the non-relativistic (NR) case, or by
\begin{equation}\label{30}
f(x_{F})\simeq \frac{1}{4}x_{F}^{4}\left(1 + \frac{1}{x_{F}}^{2}+...\right),  x_{F}\gg 1
\end{equation}
in the extreme relativistic (ER) case. The average force exerted
by the fermions a distance $r$ from the center of the star is
\cite{WD}
\begin{equation}\label{31}
<F_{0}> = \frac{\partial E_{0}}{\partial r}\simeq
\frac{4m^{4}c^{5}}{\pi\hbar^{3}}r^{2}f( x_{F}) .
\end{equation}
From (\ref{31}) and the expression
\begin{equation}\label{32}
N( r) = \frac{4}{9\pi\hbar^{3}}r^{3}{p_{F}}^{3},
\end{equation}
that gives the number of fermions in the ground state at $ r$, one
can estimate the average acceleration per fermion as a function of
$ r$
\begin{equation}\label{33}
<a(r)> = \frac{9c^2}{{x_{F}}^{3}r}f(x_{F}).
\end{equation}
By using (\ref{29}) and (\ref{30}) one finds to second order
\begin{eqnarray}\label{34}
<a(r)>_{NR}&=& \frac{3c^{2}}{r}\left(1+\frac{3}{10}x_{F}^{2}\right)\\ \nonumber
<a(r)>_{ER}&=&\frac{9c^{2}}{4r}\left(x_{F}+\frac{1}{x_{F}}\right).
\end{eqnarray}
The MA limit (\ref{8}) applied to $<a(r)>$ now yields
\begin{eqnarray}\label{35}
r\geq (r_{0})_{NR}&\equiv& \frac{3\lambda}{4\pi}\\ \nonumber r\geq
(r_{0})_{ER}&\equiv&\frac{9}{16\pi}\frac{\lambda p_{F}}{mc} ,
\end{eqnarray}
where $\lambda\equiv\frac{h}{mc} $ is the Compton wavelength of $m$. For a white dwarf, $
N/V\simeq 4.6\times 10^{29}  cm^{-3}$ gives $ (r_{0})_{NR}\simeq 5.8\times 10^{-11} cm$
and $ (r_{0})_{ER}\simeq 4\times 10^{-11} cm$.

 In order to have at least one state with particles reaching MA values, one must
have
\begin{eqnarray}\label{36}
Q\left((r_{0})_{NR}\right)&=&\frac{4}{9\pi}\left((r_{0})_{NR}\frac{p_{F}}{\hbar}\right)^{3}\sim
1 \\ \nonumber
Q\left((r_{0})_{ER}\right)&=&\frac{4}{9\pi}\left((r_{0})_{ER}\frac{p_{F}}{\hbar}\right)^{3}\sim
1 .
\end{eqnarray}
In the case of a typical white dwarf, the first of (\ref{36}) gives $ (N/V)_{NR}\sim
1.2\times 10^{30} cm^{-3}$ and the second one $(N/V)_{ER}\sim 1.3\times 10^{30} cm^{-3}$.
On the other hand, the condition $x_{F}\ll 1$ requires $ (N/V)_{NR}\ll 6\times 10^{29}
cm^{-3}$, whereas $x_{F}\gg 1$ yields $(N/V)_{ER}\gg 6\times 10^{29} cm^{-3}$. It
therefore follows that the NR approximation does not lead to electron densities
sufficient to produce states with MA electrons. The possibility to have states with MA
electrons is not ruled out entirely in the ER case.

The outlook is however different if one starts from conditions
that do not lead necessarily to the formation of canonical white
dwarfs or neutron stars \cite{WD}. Pressure consists, in fact, of
two terms. The first term represents the pressure exerted by the
small fraction of particles that can reach accelerations
comparable with MA. It is given by
\begin{equation}\label{37}
P_{MA}= \frac{2m^{2}c^{3}}{\hbar}\frac{Q(r_{0})}{4\pi r_{0}^{2}} .
\end{equation}
The second part is the contribution of those fermions in the gas ground state that can
not achieve MA
\begin{equation}\label{38}
-\frac{\partial E_{0}^{\prime}}{\partial V}= -\frac{\partial}{\partial
V}\left(2\tilde{\gamma} V \int_{r_{0}}^{\infty}d\epsilon
\frac{\epsilon^{3/2}}{e^{\alpha+\beta}+1}\right) ,
\end{equation}
where $ \tilde{\gamma} \equiv
\frac{\left(2m\right)^{3/2}}{\left(2\pi\right)^{2}\hbar^{3}}$ and $
V=\frac{4}{3}\pi\left(R^{3}-r_{0}^{3}\right)$. Since $r_{0}$ is small, one can write
\begin{equation}\label{39}
\frac{\partial E_{0}^{\prime}}{\partial V}\simeq \frac{\partial E_{0}}{\partial V}\sim
\frac{4K\tilde{M}^{5/3}}{5\tilde{R}^{5}} ,
\end{equation}
where $ \tilde{M}\equiv \frac{9\pi M}{8m_{p}}$, $ \tilde{R}\equiv \frac{2\pi
R}{\lambda}$, $K\equiv \frac{m^{4}c^{5}}{12\pi^{2}\hbar^{3}}$ and $m_{p}$ is the mass of
the proton. In the non relativistic case, the total pressure is obtained by adding
(\ref{37}) to (\ref{39}) and using the appropriate expression for $Q(r_{0})$ in
(\ref{36}). The hydrostatic equilibrium condition is
\begin{equation}\label{40}
\frac{8\pi m
c^{2}}{3\lambda^{3}}\frac{\tilde{M}}{\tilde{R}^{3}}+\frac{4K}{5}\frac{\tilde{M}^{5/3}}
{\tilde{R}^{5}}=
K^{\prime}\frac{\tilde{M}^{2}}{\tilde{R}^{4}},
\end{equation}
where $K^{\prime}\equiv \frac{4\alpha G \pi}{\lambda^{4}}\frac{64 m_{p}^{2}}{81}$ and
$\alpha \approx 1$ is a factor that reflects the details of the model used to describe
the hydrostatic equilibrium of the star. It is generally assumed that the configurations
taken by the star are polytropes \cite{shap}. Solving (\ref{40}) with respect to $
\tilde{R}$ one finds
\begin{equation}\label{41}
\tilde{R}=\frac{\tilde{M} \tilde{M_{0}}^{-2/3}}{8}\left(1 \mp \sqrt{1-\frac{64}{5}
\left(\frac{\tilde{M}_{0}}{\tilde{M}}\right)^{4/3}} \right) ,
\end{equation}
where $ \tilde{M}_{0}\equiv \left(K/K^{\prime}\right)^{3/2}=
\left(\frac{27\pi\hbar c}{64\alpha G
m_{p}^{2}}\right)^{\frac{3}{2}}\simeq
\frac{9\pi}{8m_{p}}M_{\odot}$. Solutions (\ref{41}) will be
designated by $ \tilde{R}_{-}$ and $ \tilde{R}_{+}$. They are real
if $ M\geq \left(\frac{64}{5}\right)^{3/4}M_{0}\sim 6.8
M_{\odot}$. This is a new stability condition. At the reality
point the radius of the star already is twice that of a canonical
white dwarf. This situation persists for the solution $ R_{-}$ up
to mass values of the order of $ \sim 10 M_{\odot}$. The radii of
the $ R_{+}$ solutions increase steadily. The corresponding
electron densities, calculated from $ N/V = \frac{3M}{8\pi
m_{p}R^{3}}$, are $\left(N/V\right)_{-}
> \frac{8\pi}{3}\left(\frac{8}{\lambda}\right)^{3}\left(M_{0}/M\right)^{2}\sim 6.6\times
10^{30}cm^{-3}$ and $\left(N/V\right)_{+}< 6.6\times
10^{30}cm^{-3}$. NR stars of the $ R_{-}$ type thus appear to be
more compact than those of the $ R_{+}$ class. The electron
density for $R_{+}$ is still compatible with that of a canonical
NR white dwarf. Equation (\ref{40}) can also be written in the
form
\begin{equation}\label{42}
\tilde{M}^{1/3} \tilde{R}= \frac{4}{5}\tilde{M}_{0}^{2/3}\left(1+\frac{10\pi m c^{2}}{3
\lambda^{3}K}\frac{\tilde{R}^{2}}{\tilde{M}^{2/3}}\right) ,
\end{equation}
where the second term on the r.h.s. represents the MA contribution to the usual M-R
relation for NR white dwarfs. This contribution can be neglected when $
\frac{\tilde{R}}{\tilde{M}^{1/3}}< 1/\sqrt{5}$ which requires $ N/V > \frac{8\pi
5^{3/2}}{3 \lambda^{3}} \sim 6.6 \times 10^{29} cm^{-3}$. This condition and the usual
M-R relation, $ \tilde{M}^{1/3}\tilde{R}=\frac{4\tilde{M}_{0}^{2/3}}{5} $, are compatible
if $ R < \frac{\lambda}{\pi 5^{3/4}}\left(\frac{9\pi M_{0}}{8m_{p}}\right)^{1/3}$, which
leads to the density $N/V
>\frac{5^{9/4}\pi}{3\lambda^{3}}\left(M/M_{0}\right)\sim 2.7\times 10^{30}(M/M_{0})
cm^{-3}$.

Similarly, one can calculate the MA contribution to the M-R relation for ER white dwarfs.
The equation is
\begin{equation}\label{43}
\frac{m^{4}c^{5}}{4 \pi^{2}\hbar^{3}}+K
\left(\frac{\tilde{M}^{4/3}}{\tilde{R}^{4}}-\frac{\tilde{M}^{2/3}}{\tilde{R}^{2}}\right)
=K^{\prime}\frac{\tilde{M}^{2}}{\tilde{R}^{4}} ,
\end{equation}
where the first term on the l.h.s. represents the MA contribution. From (\ref{43}) one
gets
\begin{equation}\label{44}
\tilde{R}=\tilde{M}^{1/3}\sqrt{4-\left(\frac{\tilde{M}}{\tilde{M}_{0}}\right)^{2/3}} ,
\end{equation}
and the stability condition becomes $ M\leq 8 M_{0}\sim 8
M_{\odot}$. These are very compact objects. For the electron
densities determined, the star can still be called a white dwarf.
One also finds that for $N/V\sim 3\times 10^{30}cm^{-3}$ the
number of MA states is only $
Q(r_{0})=\frac{9\lambda^{3}}{16\pi^{2}}\frac{N}{V}\simeq
2.2\frac{M}{M_{\odot}}$. A few MA electrons could therefore be
present at this density. However, interactions involving electrons
and protons at short distances may occur before even this small
number of electrons reaches the MA.

Analogous conclusions also apply to neutron stars, with minor changes if these can be
treated as Newtonian polytropes. This approximation may only be permissible, however, for
low-density stars \cite{shap}. One finds in particular $ Q(r_{0})=\frac{9
\lambda_{n}^{3}}{16\pi^{2}}\frac{N}{V}\simeq 4.5 \frac{M}{M_{\odot}}$. The presence of a
few MA neutrons is therefore allowed in this case.

\section{Conclusions}

A limit on the proper acceleration of particles can be obtained
from the uncertainty relations in the following way. Ehrenfest
theorem (\ref{6}) is first applied to a particle's acceleration $
\vec{a}$ in the particle's instantaneous rest frame. The latter is
then transformed to a Lorentz frame of instantaneous velocity $
\vec{v}$. In any other Lorentz frame the resulting acceleration is
$ a' \leq a_{p}$.  The absolute value of the proper acceleration
satisfies (\ref{9}). No counterexamples are known to the validity
of (\ref{9}). For this reason (\ref{9}) has at times been elevated
to the status of a principle. It would be only befitting to call
it the Caianiello principle.

In most instances the value of MA is so high that it defies direct
observation. Nonetheless the role of MA as a universal regulator
must not be discounted. It is an intrinsic, first quantization
limit that preserves the continuity of space-time and does not
require the introduction of a fundamental length, or of arbitrary
cutoffs. The challenge is to find situations where MA affects the
physics of a system in ways that can be observed.

Though the existence of a MA is intimately linked to the validity
of Ehrenfest theorem, it is not entirely subordinate to it and the
limit itself may depend on the dynamical characteristics of the
particular system considered. This is the case with
superconductors of the first kind which are macroscopic,
non-relativistic quantum systems with velocities that satisfy the
inequality $ \Delta E \Delta v \ll m c^{3}$. Superfluid particles
in static conditions are known to resist acceleration. The MA
constraints (\ref{11}) and (\ref{12}) lead to a limit on the value
of the electric field at the surface of the superconductor that
agrees with the value (\ref{19}) obtained from London's equations
and with known experimental results. The MA limit in this case is
only $ \sim 10 $ times larger than (\ref{**}). If the pairing
condition is satisfied, then superfluids in static conditions obey
a dynamics for which a MA {\it exists} and differs from $
{\cal{A}}_{m}$, as anticipated.

In Section (III) two high-energy particles, typically leptons,
produce a third particle at rest. The MA limit applied to the
process leads to the constraint (\ref{21}) on the width of the
particle produced. The limit is perfectly consistent with
available experimental results. When the end product is $ Z^{0}$,
the acceleration is $ a_{r}= \frac{2mc}{\hbar m_{D}}\Gamma
(AB)\sim 2.8 \times 10^{26}\,cm\, s^{-2}$. Even at these energies
the value of the acceleration is only a factor $ \sim 6\times
10^{-6}$ that of the MA for $ m_{r}$. Equation (\ref{20}) and
current estimates of $ \Gamma (ZZ\rightarrow H^{0})$ can also be
used to derive upper and lower limits (\ref{00}) on the mass of
the Higgs boson.

The last physical situation considered regards matter in the
interior of white dwarfs and neutron stars. For canonical white
dwarfs, the possibility that states exist with MA electrons can be
ruled out in the NR case, but not so for ER stars. On the other
hand, the mere presence of a few MA electrons alters the stability
conditions of the white dwarf drastically. Equations (\ref{42})
and (\ref{44}) represent, in fact, {\it new stability conditions}.
The same conclusions also apply to NR neutron stars, with
limitations, however, on the choice of the equation of state. In
the collapse of stars with masses larger than the Chandrasekhar
and Oppenheimer-Volkoff limits from white dwarfs to neutron stars,
to more compact objects, conditions favorable to the formation of
states with MA fermions may occur before competing processes take
place.


\begin{thebibliography}{99}
\bibitem{cai2} E.R. Caianiello, Lett. Nuovo Cimento {\bf 41}, 370 (1984). See also E. R.
Caianiello, Rivista Nuovo Cimento {\bf 15}, No. 4 (1992).
\bibitem{wood} W.R. Wood, G. Papini, Y.Q. Cai, Nuovo Cimento {\bf 104 B}, 361 and errata
727 (1989).
\bibitem{land} L.D. Landau and E.M. Lifshitz, {\it Quantum Mechanics}, third edition (Pergamon
Press, New York, 1977), pp.27 and 49.
\bibitem{mess} Albert Messiah, {\it Quantum Mechanics}, Vol. I
(North-Holland, Amsterdam, 1961), Chapters IV.10 and VIII.13.
\bibitem{bal} Roger Balian, {\it From Microphysics to Macrophysics.
Methods and Applications of Statistical Physics, Vol. I} (Springer-Verlag, Berlin, 1991).
\bibitem{steph} G. Stephenson and C.W. Kilmister, {\it Special Relativity for Physicists}
(Longmans, London, 1965).
\bibitem{marg} Norman Margolus and Lev B. Levitin, Physica D {\bf
120}, 188 (1998).
\bibitem{beck} Jacob D. Beckenstein, Phys. Rev. Lett. {\bf 46},
623 (1981).
\bibitem{an} J. Anandan and Y. Aharonov, Phys. Rev. Lett. {\bf
65}, 1697 (1990).
\bibitem{pap02} G. Papini, Nuovo Cimento {\bf 117 B}, 1325 (2002).
\bibitem{heg} G.C. Hegerfeldt, Phys. Rev. D {\bf 10}, 3320 (1974).
\bibitem{mthw} C.W. Misner, K.S. Thorne, J.A. Wheeler, {\it Gravitation} (Freeman, S.
Francisco, 1973).
\bibitem{das} A. Das, J. Math. Phys. {\bf 21}, 1506 (1980).
\bibitem{gasp1} M. Gasperini, Astroph. Space Sci. {\bf 138}, 387 (1987).
\bibitem{toller} M. Toller, Nuovo Cimento {\bf 102 B}, 261(1988); Int. J. Theor. Phys.
{\bf 29}, 963 (1990); Phys. Lett. B {\bf 256}, 215 (1991).
\bibitem{paren} R. Parentani, R. Potting, Phys. Rev. Lett. {\bf 63}, 945 (1989).
\bibitem{vora} P. Voracek, Astroph. Space Sci. {\bf 159}, 181 (1989).
\bibitem{mash} B. Mashhoon, Phys Lett. A {\bf 122}, 67, 299 (1987);{\bf 143}, 176 (1990);
{\bf 145}, 147 (1990).
\bibitem{venzo} V. de Sabbata, C. Sivaram, Astroph. Space Sci. {\bf 176}, 145 (1991);
{\it Spin and Torsion in Gravitation} (World Scientific, Singapore, 1994).
\bibitem{falla} D.F. Falla, P.T. Landsberg, Nuovo Cimento {\bf 106 B}, 669 (1991).
\bibitem{pati} A.K. Pati, Nuovo Cimento {\bf 107 B}, 895 (1992); Europhys. Lett. {\bf
18}, 285 (1992).
\bibitem{nester} V.V. Nesterenko, A. Feoli, G. Lambiase and G. Scarpetta, Phys. Rev. D
{\bf 60}, 065001 (1999).
\bibitem{hoof} G. 't Hooft, Nucl. Phys. B {\bf 256}, 727 (1985).
\bibitem{suss} L. Susskind, J. Uglum, Phys. Rev. D {\bf 50}, 2700 (1994).
\bibitem{mcgui} M. McGuigan, Phys. Rev. D {\bf 50}, 5225 (1994).
\bibitem{cai4} E.R. Caianiello, Nuovo Cimento {\bf 59 B}, 350 (1980); G. Scarpetta, Lett.
Nuovo Cimento {\bf 51}, 51 (1954); E.R. Caianiello, G. Scarpetta, G. Marmo, Nuovo Cimento
{\bf 86 A}, 337 (1985); E.R. Caianiello, A. Feoli, M. Gasperini, G. Scarpetta, Int. J.
Theor. Phys. {\bf 29}, 131 (1990).
\bibitem{pap} G. Papini and W.R. Wood, Phys. Lett. A {\bf 170}, 409 (1992); W.R. Wood and
G. Papini, Phys. Rev. D {\bf 45}, 3617 (1992); Found. Phys. Lett. {\bf 6}, 409 (1993).
\bibitem{sanchez} N. Sanchez, G. Veneziano, Nucl. Phys. B {\bf 333}, 253 (1990).
\bibitem{gasp2} M. Gasperini, N. Sanchez, G. Veneziano, Nucl. Phys. B {\bf 364}, 365
(1991); Int. J. Mod. Phys. A {\bf 6}, 3853 (1991).
\bibitem{gasp3} M. Gasperini, Phys. Lett. B {\bf 258}, 70 (1991); Gen. Rel. Grav. {\bf
24}, 219 (1992).
\bibitem{fro} V.P. Frolov, N. Sanchez, Nucl. Phys. B {\bf 349}, 815 (1991).
\bibitem{pap1} G. Papini, A. Feoli, G. Scarpetta, Phys. Lett. A {\bf 202}, 50 (1995); G.
Lambiase, G. Papini, G. Scarpetta, Phys. Lett. A {\bf 244}, 349 (1998); Nuovo Cimento
{\bf 114 B}, 189 (1999); A. Feoli, G. Lambiase, G. Papini, G. Scarpetta Phys. Lett. A
{\bf 263}, 147 (1999); S. Capozziello, A. Feoli, G. Lambiase, G. Papini, G. Scarpetta,
Phys. Lett. A {\bf 268}, 247 (2000); V. Bozza, A. Feoli, G. Papini, G. Scarpetta, Phys.
Lett. A {\bf 271}, 35 (2000); {\bf 279}, 163 (2001); {\bf 283}, 53 (2001).
\bibitem{cai5} E.R. Caianiello, M. Gasperini, G. Scarpetta, Nuovo Cimento {\bf 105 B},
259 (1990); E.R. Caianiello, M. Gasperini, E. Predazzi, G. Scarpetta, Phys. Lett. A {\bf
132}, 83 (1988).
\bibitem{cai1} E.R. Caianiello, Lett. Nuovo Cimento {\bf 32}, 65 (1981).
\bibitem{degennes} P.G. De Gennes, {\it Superfluidity of Metals and Alloys} (W.A. Benjamin, New
York, 1966).
\bibitem{til} D.R. Tilley and J. Tilley, {\it Superfluidity and Superconductivity}, third edition
(Adam Hilger, Bristol, 1990).
\bibitem{london} F. London, {\it Superfluids, Vol. I} (Dover Publications, New York,
1961).
\bibitem{bok} J. Bok and J. Klein, Phys. Rev. Lett. {\bf 20}, 660 (1968).
\bibitem{weber} J. Weber, Nuovo Cimento {\bf 109 B}, 855 (1994).
\bibitem{renton} See, for instance: Peter Renton, {\it Electroweak Interactions} (Cambridge
University Press, Cambridge, 1990).
\bibitem{kuwata} S. Kuwata, Nuovo Cimento {\bf 111 B}, 893, (1996).
\bibitem{WD} G. Papini, Physics Lett. A {\bf 305}, 359 (2002).
\bibitem{shap} Stuart L. Shapiro and Saul A. Teukolsky, {\it Black Holes, White Dwarfs
and Neutron Stars} (John Wiley and Sons, New York, 1983).



\end{thebibliography}
\end{document}